\documentclass[12pt]{article}
\usepackage{latexsym,amssymb,amsfonts,amsmath}
\usepackage{graphicx}
\usepackage{color}

\newlength{\extraspace}
\newlength{\extraspaces}
\newcommand{\be}{\begin{equation}
\addtolength{\abovedisplayskip}{\extraspaces}
\addtolength{\belowdisplayskip}{\extraspaces}
\addtolength{\abovedisplayshortskip}{\extraspace}
\addtolength{\belowdisplayshortskip}{\extraspace}}
\newcommand{\ee}{\end{equation}}

\newcommand{\bea}{\begin{eqnarray}
\addtolength{\abovedisplayskip}{\extraspaces}
\addtolength{\belowdisplayskip}{\extraspaces}
\addtolength{\abovedisplayshortskip}{\extraspace}
\addtolength{\belowdisplayshortskip}{\extraspace}}
\newcommand{\eea}{\end{eqnarray}}

\DeclareMathOperator{\Tr}{Tr} 
\DeclareMathOperator{\diag}{diag} 

\newcommand{\Fpq}{f_{a}}

\newcommand{\pmud}{\partial_{\mu}}
\newcommand{\pmuu}{\partial^{\mu}}

\newcommand{\dd}{\mathrm{d}}

\newcommand{\dgg}{^\dagger}

\begin{document}

\addtolength{\baselineskip}{.8mm}

{\thispagestyle{empty}

\mbox{}                \hfill Revised: July 2020 \hspace{1cm}\\

\begin{center}
{\Large\bf QCD axion and topological susceptibility}\\
{\Large\bf in chiral effective Lagrangian models}\\
{\Large\bf at finite temperature}\\
\vspace*{1.0cm}
{\large
Salvatore Bottaro\footnote{E.mail: salvatore.bottaro@sns.it}
}\\
\vspace*{0.5cm}{\normalsize
{\textit{Scuola Normale Superiore and INFN, Sezione di Pisa,\\
Piazza dei Cavalieri 7, 56126, Pisa, Italy}}}\\
\vspace*{0.5cm}
{\large
Enrico Meggiolaro\footnote{E-mail: enrico.meggiolaro@unipi.it}
}\\
\vspace*{0.5cm}{\normalsize
{\textit{Dipartimento di Fisica, Universit\`a di Pisa,
and INFN, Sezione di Pisa,\\
Largo Pontecorvo 3, I-56127 Pisa, Italy}}}\\
\vspace*{2cm}{\large \bf Abstract}
\end{center}

\noindent
In this work we compute the axion mass and, from this (exploiting a well-known relation), we also derive an expression for the QCD topological susceptibility in the finite-temperature case, both below and above the chiral phase transition at $T_c$, making use of a chiral effective Lagrangian model which includes the axion, the scalar and pseudoscalar mesons and implements the $U(1)$ axial anomaly of the fundamental theory. We also provide a numerical estimate of the topological susceptibility at $T=0$ (in the physical case of three light quark flavors) and discuss the question of the temperature and quark-mass dependence of the topological susceptibility in the high-temperature regime.

}
\newpage

\section{Introduction}
\label{sec1}

Among the possible solutions of the so-called ``strong-$CP$ problem'' (that is, the experimental absence of $CP$ violations in the strong-interaction sector), the most appealing is surely the one proposed by Peccei and Quinn (PQ) in 1977 \cite{PQ1977} and developed by Weinberg and Wilczek in 1978 \cite{Weinberg1978,Wilczek1978}.
The key idea (see also Ref. \cite{Peccei2008} for a recent review) is to extend the Standard Model by adding a new pseudoscalar particle, called ``axion'', in such a way that there is a new $U(1)$ global symmetry, referred to as $U(1)_{PQ}$, which is both spontaneously broken at a scale $f_a$ and anomalous (i.e., broken by quantum effects), with the related current $J^{\mu}_{PQ}$ satisfying the relation
\begin{equation}
\label{anomaly}
\partial_{\mu}J^{\mu}_{PQ}=a_{PQ}Q ,
\end{equation} 
where
$Q = \frac{g^2}{64\pi^2}\varepsilon^{\mu\nu\rho\sigma}F^a_{\mu\nu}F^a_{\rho\sigma}$ is the \emph{topological charge density} and $a_{PQ}$ is the so-called \emph{color anomaly} parameter.
The most general Lagrangian describing the QCD degrees of freedom $\Psi$ and the axion field $S_a$ has the following form:
\begin{equation}
\label{lag_qcdax}
\mathcal{L} = \mathcal{L}_{QCD} +\frac{1}{2}\pmud S_a \pmuu S_a+\mathcal{L}_{int}[\pmud S_a,\Psi]-\frac{a_{PQ}}{\Fpq}S_a Q ,
\end{equation}
where the term $\mathcal{L}_{int}[\partial_{\mu}S_a,\Psi]$ describes the interactions between the axion and the quark fields and it is strongly model dependent. Under $U(1)_{PQ}$ the axion field $S_a$ transforms nonlinearly as
\begin{equation} \label{shiftaxion}
U(1)_{PQ}:~~~~ S_a\rightarrow S_a'= S_a + \gamma f_{a} ,
\end{equation}
so that the first three terms in \eqref{lag_qcdax} are left invariant, while the last one reproduces the correct anomaly of \eqref{anomaly}.
By virtue of this extra $U(1)_{PQ}$ symmetry, $CP$ comes out to be dynamically conserved in this model.

Moreover, it is well known that the $U(1)$ axial symmetry of QCD with $n_l$ light quark flavors (taken to be massless in the ideal \textit{chiral limit}; the physically relevant cases being $n_l=2$, with the quarks \textit{up} and \textit{down}, and $n_l=3$, including also the \textit{strange} quark),
\begin{equation} \label{U(1)_A}
U(1)_A:~~~~ q_i \rightarrow q'_i = e^{i\beta\gamma_5}q_i,~~~ i=1,\ldots,n_l ,
\end{equation}
is also anomalous, with the related $U(1)$ axial current $J^{\mu}_{5}=\bar{q}\gamma^{\mu}\gamma_{5}q$ satisfying the relation $\partial_{\mu}J^{\mu}_{5}=2n_lQ$.
Therefore, we find that the $U(1)_A \otimes U(1)_{PQ}$ transformations with the parameters $\beta$ and $\gamma$ satisfying the constraint $2n_l\beta + a_{PQ} \gamma = 0$ form a $U(1)$ subgroup which is spontaneously broken but anomaly-free (in the chiral limit): as a consequence, a new (pseudo-)Nambu-Goldstone boson appears in the spectrum, the axion.

Indeed, the Lagrangian \eqref{lag_qcdax} is already sufficient to derive an important relation (first introduced in Refs. \cite{SVZ1980,PWW1983,AS1983,DF1983}) between the axion mass and the \emph{topological susceptibility} of QCD, defined as $\chi_{QCD} = -i\int\dd^4x\langle T\{Q(x)Q(0)\}\rangle_{QCD}$, namely,
\begin{equation}
\label{axmass}
m^2_{axion} \simeq \frac{a_{PQ}^2}{\Fpq^2}\chi_{QCD} ,
\end{equation}
which is valid at the leading order in $1/\Fpq$, assuming that $\Fpq$ is much larger than the QCD scale ($\Fpq \gg \Lambda_{QCD}$).
Indeed, this assumption is phenomenologically well established, since at present (see, for example, Refs. \cite{bounds_a,bounds_b}) astrophysical and cosmological considerations imply the following bounds on the $U(1)_{PQ}$ breaking scale $\Fpq$ (or, better, on $\Fpq/a_{PQ}$, but $a_{PQ} \sim \mathcal{O}(1)$ for the more realistic axion models \cite{DiLuzio2017}): $10^9~ \text{GeV} \lesssim \Fpq \lesssim 10^{17}~ \text{GeV}$.

In this paper we shall consider the relation \eqref{axmass} in the theory at a finite temperature $T$.
(See also Ref. \cite{Farias2019} for a recent investigation of the effects of a hot and magnetized medium on the axion mass and the QCD topological susceptibility, making use of the Nambu-Jona-Lasinio effective model.)
It is well known (mainly by lattice simulations \cite{HotQCD}) that, at temperatures above a certain (pseudo)critical temperature $T_c \approx 150$ MeV, thermal fluctuations break up the chiral condensate $\langle \bar{q} q \rangle$, causing the complete restoration of the $SU(n_l)_L\otimes SU(n_l)_R$ chiral symmetry of QCD with $n_l$ light quarks ($n_l=2$ and $n_l=3$ being the physically relevant cases): this leads to a phase transition called ``chiral transition''. For what concerns, instead, the $U(1)$ axial symmetry, the nonzero contribution to the anomaly provided by the instanton gas at high temperatures \cite{GPY1981} should imply that it is always broken, also for $T>T_c$. (However, the real magnitude of its breaking and its possible \emph{effective} restoration at some temperature above $T_c$ are still important debated questions in hadronic physics.)

In this work we shall compute the axion mass and therefore, exploiting the relation \eqref{axmass}, we shall also derive an expression for the QCD topological susceptibility in the finite-temperature case, both below and above the chiral phase transition at $T_c$, making use of a chiral effective Lagrangian model, the so-called ``interpolating model'', which includes the axion, the scalar and pseudoscalar mesons and implements the $U(1)$ axial anomaly of the fundamental theory.
The inclusion of the axion in a low-energy effective Lagrangian model of QCD is, of course, fully justified, since, being $\Fpq \gg \Lambda_{QCD}$, the axion is an extremely light degree of freedom (its mass being smaller than about $0.01$ eV).
The choice of the interpolating model, described in detail in the next section, is due to its ``regularity'' around the chiral phase transition (i.e., it is well defined also above $T_c$) and to the fact that the other known effective Lagrangian models (and the corresponding results for the axion mass and the QCD topological susceptibility, both below and above $T_c$) can be obtained by taking proper formal limits of the interpolating model (and its results), as already noticed in Ref. \cite{EM2019} (for the chiral effective Lagrangian models without the axion). The advantages of this approach of computing $\chi_{QCD}$, as we shall see, is that, being the axion a pseudoscalar particle and $CP$ now an exact symmetry, there can be no mixing with the scalar degrees of freedom of the effective model (which must be included if we want to perform our analysis also at temperatures around and above the chiral phase transition), so that the problem reduces to finding the lightest particle (with a mass vanishing as $1/\Fpq$ when $\Fpq \to \infty$) among the pseudoscalar degrees of freedom.

The plan of the paper is the following. In Sec. \ref{sec2} we shall present the (linearized) interpolating model with the inclusion of the axion and we shall discuss its relation with other known effective models.
In Sec. \ref{sec3} we shall compute the axion mass and thus the topological susceptibility at finite temperature, both below and above the chiral transition, using the interpolating model: from this, using the correspondence relations found in Sec. \ref{sec2}, we shall also derive the expression of the topological susceptibility for other known effective Lagrangian models.
In the Appendix, we shall also give a numerical evaluation of the expressions for the topological susceptibility at zero temperature in the physical case $n_l=3$. Finally, in Sec. \ref{sec4} we shall briefly summarize the results obtained in this paper, giving some prospects and conclusions.

\section{The interpolating model with the axion}
\label{sec2}

The effective Lagrangian model that we shall consider (originally proposed in Ref. \cite{EM1994} and elaborated on in Refs. \cite{MM2003,EM2011,MM2013}) is a generalization of the model proposed (in the context of the large-$N_c$ expansion) by Witten, Di Vecchia, Veneziano, \emph{et al.} \cite{WDV1,WDV2,WDV3a,WDV3b,WDV3c,WDV3d} (that, following the notation introduced in Refs. \cite{EM2019,LM2018}, will be denoted for brevity as the ``WDV model'').
Following Refs. \cite{EM2019,LM2018}, we shall call it the ``interpolating model'' (IM), because (in a sense which will be recalled below) it approximately ``interpolates'' between the WDV model at $T=0$ and the so-called ``extended linear sigma (EL$_\sigma$) model'' for $T>T_c$.
The EL$_\sigma$ model was originally proposed in Ref. \cite{ELSM1a,ELSM1b,ELSM1c} to study the chiral dynamics at $T=0$, and later used as an effective model to study the chiral-symmetry restoration at nonzero temperature \cite{PW1984,ELSMfiniteT_1a,ELSMfiniteT_1b,ELSMfiniteT_2a,ELSMfiniteT_2b,ELSMfiniteT_2c}: according to 't Hooft (see Refs. \cite{ELSM2,ELSM3} and references therein), it reproduces, in terms of an effective theory, the $U(1)$ axial breaking caused by instantons in the fundamental theory.\footnote{We recall here, however, the criticism by Christos \cite{Christos1984} (see also Refs. \cite{WDV1,WDV2}), according to which the determinantal interaction term in this effective model does not correctly reproduce the $U(1)$ axial anomaly of the fundamental theory.}

In the interpolating model the $U(1)$ axial anomaly is implemented, as in the WDV model, by properly introducing the topological charge density $Q$ as an auxiliary field, so that it satisfies the correct transformation property under the chiral group (and is consistent with the large-$N_c$ expansion).\footnote{However, we must recall here that also the particular way of implementing the $U(1)$ axial anomaly in the WDV model, by means of a logarithmic interaction term [as in Eqs. \eqref{IM+Q} and \eqref{IM_potential} below], was criticized by 't Hooft in Ref. \cite{ELSM2}.}
Moreover, it also assumes that there is another $U(1)$-axial-breaking condensate (in addition to the usual quark-antiquark chiral condensate $\langle \bar{q}q \rangle$), having the form $C_{U(1)} = \langle {\cal O}_{U(1)} \rangle$,
where, for a theory with $n_l$ light quark flavors, ${\cal O}_{U(1)}$ is a
$2n_l$-quark local operator that has the chiral transformation properties of
\cite{tHooft1976,KM1970,Kunihiro2009}
${\cal O}_{U(1)} \sim \displaystyle{{\det_{st}}(\bar{q}_{sR}q_{tL})
+ {\det_{st}}(\bar{q}_{sL}q_{tR}) }$,
where $s,t = 1,\ldots,n_l$ are flavor indices.\footnote{The explicit form of the condensate (including the color indices) for the cases $n_l=2$ and $n_l=3$ is discussed in detail in Appendix A of Ref. \cite{EM2011}.}
The effective Lagrangian of the interpolating model is written in terms
of the topological charge density $Q$, the mesonic field
$U_{ij} \sim \bar{q}_{jR} q_{iL}$ (up to a multiplicative constant),
and the new field variable $X \sim {\det} \left( \bar{q}_{sR} q_{tL} \right)$
(up to a multiplicative constant), associated with the $U(1)$ axial condensate:
\begin{equation}\label{IM+Q}
\begin{split}
\mathcal{L}_{IM}(U,&U^\dagger,X,X^\dagger,Q)
=\frac{1}{2}\Tr [\partial_\mu U \partial^\mu U^{\dagger}]
+ \frac{1}{2}\partial_\mu X \partial^\mu X^{\dagger}
-V_0(U,U^\dagger,X,X^\dagger) \\
&+ \frac{i}{2}Q \left[ \omega_1 \Tr (\log U -\log U^\dagger)
+ (1-\omega_1) (\log X -\log X^\dagger) \right] + \frac{1}{2A}Q^2 ,
\end{split}
\end{equation}
where
\begin{equation}\label{IM_potential_0}
\begin{split}
V_0(U,U^\dagger,X,X^\dagger) &= \frac{\lambda_\pi^2}{4}\Tr [(UU^{\dagger}-\rho_\pi \mathbf{I})^2] + \frac{\lambda_\pi^{'2}}{4}\left[\Tr(UU^\dagger)\right]^2 + \frac{\lambda_X^2}{4} [XX^\dagger - \rho_X]^2 \\ &- \frac{B_m}{2\sqrt{2}}\Tr\left[M(U + U^{\dagger})\right] - \frac{\kappa_1}{2\sqrt{2}}[X^\dagger \det U + X\det U^\dagger ] ,
\end{split}
\end{equation}
$M = \diag (m_1,\ldots,m_{n_l})$ being the physical (real and diagonal) quark-mass matrix.\\
As in the case of the WDV model, the auxiliary field $Q$ in \eqref{IM+Q} can be integrated out using its equation of motion, obtaining
\begin{equation}\label{IM}
\mathcal{L}_{IM}(U,U^\dagger,X,X^\dagger)=\frac{1}{2}\Tr [\partial_\mu U \partial^\mu U^{\dagger}] + \frac{1}{2}\partial_\mu X \partial^\mu X^{\dagger}
-V(U,U^\dagger , X,X^\dagger) ,
\end{equation}
where
\begin{equation}\label{IM_potential}
\begin{split}
V(U,&U^\dagger,X,X^\dagger)=V_0(U,U^\dagger,X,X^\dagger) \\
&-\frac{A}{8} \left[\omega_1 \Tr (\log U - \log U^\dagger)+ (1-\omega_1)(\log X - \log X^\dagger )\right]^2 .
\end{split}
\end{equation}
We remind the reader that the \emph{only} anomalous term in the Lagrangian \eqref{IM+Q}--\eqref{IM_potential_0} of the interpolating model is the term proportional to the topological charge density $Q$, depending on $\Tr(\log U)$ and $\log X$, i.e., after integrating out the auxiliary field $Q$, the last term (proportional to $A$) in Eq. \eqref{IM_potential}: this term has exactly the same structure of the anomalous term in the WDV model and guarantees that the Lagrangian correctly transforms under $U(1)$ axial transformations.
On the contrary, the last interaction term in Eq. \eqref{IM_potential_0}, proportional to $X^\dagger \det U + X \det U^\dagger$, while being very similar to the interaction term of the EL$_\sigma$ model, is \emph{not} anomalous, but (since $X$ transforms exactly as $\det U$ under a chiral group transformation) it is invariant under the entire chiral group $U(n_l) \otimes U(n_l)$.

All the parameters which appear in Eqs. \eqref{IM_potential_0} and \eqref{IM_potential} have to be considered as temperature dependent.
In particular, we recall that the parameter $\rho_\pi$ is responsible for the fate of the $SU(n_l)_L \otimes SU(n_l)_R$ chiral symmetry, which, as is well known, depends on the temperature $T$: $\rho_\pi$ will be positive, and, correspondingly, the ``vacuum expectation value'' (VEV), i.e., the thermal average, of $U$ will be different from zero in the chiral limit $M=0$, until the temperature reaches the chiral phase-transition temperature $T_c$ [$\rho_\pi(T<T_c)>0$], above which it will be negative [$\rho_\pi(T>T_c)<0$], and, correspondingly, the VEV of $U$ will vanish in the chiral limit $M=0$.\footnote{We notice here that we have identified the temperature $T_{\rho_\pi}$ at which the parameter $\rho_\pi$ is equal to zero with the chiral phase-transition temperature $T_c$: this is always correct except in the case $n_l=2$, where we have $T_{\rho_\pi} < T_c$ (see Refs. \cite{EM2019,MM2013} for a more detailed discussion).}
Similarly, the parameter $\rho_X$ plays for the $U(1)$ axial symmetry the same role the parameter $\rho_\pi$ plays for the $SU(n_l)_L \otimes SU(n_l)_R$ chiral symmetry: $\rho_X$ determines the VEV of the field $X$, which is an order parameter of the $U(1)$ axial symmetry. In order to reproduce the scenario we are interested in, that is, the scenario in which the $U(1)$ axial symmetry is \emph{not} restored for $T>T_c$, while the $SU(n_l) \otimes SU(n_l)$ chiral symmetry is restored as soon as the temperature reaches $T_c$, we must assume that, differently from $\rho_\pi$, the parameter $\rho_X$ remains positive across $T_c$, i.e., $\rho_\pi(T<T_c)>0$, $\rho_X(T<T_c)>0$, and $\rho_\pi(T>T_c)<0$, $\rho_X(T>T_c)>0$.

For what concerns the parameter $\omega_1(T)$, in order to avoid a singular behavior of the anomalous term in the potential \eqref{IM_potential} above the chiral-transition temperature $T_c$, where the VEV of the mesonic field $U$ vanishes (in the chiral limit $M=0$), we must assume that \cite{EM1994,MM2013} $\omega_1 (T\geq T_c)=0$.
(This way, indeed, the term including $\log U$ in the potential vanishes, eliminating the problem of the divergence, at least as far as the VEV of the field $X$ is different from zero or, in other words, as far as the $U(1)$ axial symmetry remains broken also above $T_c$.)

At this point we can introduce the axion in our effective Lagrangian model. If we write 
\begin{equation}
N = e^{i\frac{S_a}{\Fpq}} ,
\end{equation}
it is sufficient to add to the Lagrangian \eqref{IM} a few terms:
\begin{equation}
\mathcal{L}_{IM+axion}= \mathcal{L}_{IM}+\frac{\Fpq^2}{2}\pmud N\pmuu N\dgg+\frac{i}{2}a_{PQ}(\log N-\log N\dgg)Q .
\end{equation}
This is precisely how the axion is introduced in the WDV model \cite{DS2014}, since the anomaly is implemented in the same way, and in fact it is easy to verify that the modified Lagrangian has all the required properties described in the previous section. Finally, we can eliminate $Q$ through its equation of motion to get the final Lagrangian that we shall use throughout this paper:
\begin{equation}
\label{IM+axion}
\begin{split}
\mathcal{L}_{IM+axion}=&~\frac{1}{2}\Tr[\pmud U\pmuu U\dgg]+\frac{1}{2}\pmud X\pmuu X\dgg+\frac{\Fpq^2}{2}\pmud N\pmuu N\dgg\\
&-V(U,U\dgg,X,X\dgg,N,N^\dagger) ,
\end{split}
\end{equation}
where
\begin{equation}\label{IM+axion_potential}
\begin{split}
V(U,&U^\dagger,X,X^\dagger,N,N^\dagger)=V_0(U,U^\dagger,X,X^\dagger)\\
&-\frac{A}{8} \left[\omega_1 \Tr (\log U - \log U^\dagger) + (1-\omega_1)(\log X - \log X^\dagger ) + a_{PQ}(\log N-\log N\dgg) \right]^2 .
\end{split}
\end{equation}
Now we will clarify in which sense this model interpolates between the WDV and the EL$_\sigma$ models with the inclusion of the axion, extending what was already noticed in Ref. \cite{EM2019} for the models without the axion.
As it had been already observed in Refs. \cite{EM2011,LM2018}, the Lagrangian of the WDV model is obtained from that of the interpolating model by first fixing $\omega_1=1$ and then taking the formal limits $\lambda_X \to +\infty$ and also $\rho_X \to 0$ (so that $X \to 0$).
The same statement also applies to the models with the inclusion of the axion,
the presence of this being irrelevant for these limits, i.e.,
\begin{equation}\label{IM to WDV}
\mathcal{L}_{IM+axion}\vert_{\omega_1=1} \mathop{\longrightarrow}_{\lambda_X \to +\infty,~\rho_X \to 0} \mathcal{L}_{WDV+axion} ,
\end{equation}
where (see Ref. \cite{DS2014})
\begin{equation}
\label{WDV+axion}
\begin{split}
\mathcal{L}_{WDV+axion}=&~\frac{1}{2}\Tr[\pmud U\pmuu U\dgg]+\frac{\Fpq^2}{2}\pmud N\pmuu N\dgg -V_0(U,U\dgg)\\
&+\frac{A}{8}\left[\Tr(\log U-\log U\dgg)+a_{PQ}(\log N-\log N\dgg)\right]^2 ,
\end{split}
\end{equation}
with
\begin{equation}
V_0(U,U\dgg) = -\frac{B_m}{2\sqrt{2}}\Tr[M(U+U\dgg)]+\frac{\lambda_{\pi}^2}{4}\Tr[(UU\dgg-\rho_{\pi}\mathbf{I})^2]+\frac{\lambda^{'2}_{\pi}}{4}[\Tr(UU\dgg)]^2 .
\end{equation}
On the other side, as we have seen above, the parameter $\omega_1$ must be necessarily taken to be equal to zero above the critical temperature $T_c$, where the WDV is no more valid (because of the singular behavior of the anomalous term in the potential), and vice versa, as it was already observed in Ref. \cite{MM2013}, the interaction term $\frac{\kappa_1}{2\sqrt{2}}[X^\dagger \det U + X\det U^\dagger]$ of the interpolating model becomes very similar to the ``instantonic'' interaction term $\kappa_I [\det U + \det U^\dagger]$ of the EL$_\sigma$ model.
More precisely, it was observed in Ref. \cite{EM2019} that, by first fixing $\omega_1=0$ and then taking the formal limits $\lambda_X \to +\infty$ and $A \to \infty$ (so that, writing $X= \alpha e^{i\beta}$, one has $\alpha \to \sqrt{\rho_X}$ and $\beta \to 0$, i.e., $X \to \sqrt{\rho_X}$), the Lagrangian of the interpolating model (without the axion) reduces to the Lagrangian of the EL$_\sigma$ model with $\kappa_I=\frac{\kappa_1\sqrt{\rho_X}}{2\sqrt{2}}$ (i.e., with $\kappa_I$ proportional to the $U(1)$ axial condensate).\\
The same statement also applies to the models with the inclusion of the axion, apart from a rescaling in the Peccei-Quinn scale, i.e.,
\begin{equation}\label{IM to ELSM}
\mathcal{L}_{IM+axion}\vert_{\omega_1=0} \mathop{\longrightarrow}_{\lambda_X \to +\infty,~A \to +\infty} \mathcal{L}_{EL_\sigma+axion}\vert_{\kappa_I=\frac{\kappa_1\sqrt{\rho_X}}{2\sqrt{2}},~\Fpq \to \tilde{\Fpq} = \sqrt{\Fpq^2 + a_{PQ}^2 \rho_X}} ,
\end{equation}
where
\begin{equation}
\label{ELSM+axion}
\begin{split}
\mathcal{L}_{EL_\sigma+axion}=&~\frac{1}{2}\Tr[\pmud U\pmuu U\dgg]+\frac{\Fpq^2}{2}\pmud N\pmuu N\dgg-V_0(U,U\dgg)\\
&+\kappa_I[N^{a_{PQ}}\det U + (N^{\dagger})^{a_{PQ}} \det U^{\dagger}] .
\end{split}
\end{equation}
In fact, taking the formal limits $\lambda_X \to +\infty$ and $A \to \infty$ in the interpolating model with the axion, one now gets $X= \alpha e^{i\beta} \to \sqrt{\rho_X} e^{-ia_{PQ}\frac{S_a}{f_a}}$. As a first consequence, this leads to an additional term coming from the kinetic term of the field $X$,
which renormalizes the preexisting axion kinetic term to give
\begin{equation}
\frac{1}{2}\pmud X\pmuu X\dgg + \frac{1}{2}\pmud S_a \pmuu S_a \rightarrow \frac{1}{2}\left(1+a_{PQ}^2\frac{\rho_X}{\Fpq^2}\right)\pmud S_a \pmuu S_a ,
\end{equation}
so that we have to rescale the axion field in order for it to be canonically normalized:
\begin{equation}
S_a\rightarrow\tilde{S_a}=\left(1+a_{PQ}^2\frac{\rho_X}{\Fpq^2}\right)^{1/2} S_a = \frac{\tilde{\Fpq}}{\Fpq} S_a ,~~~~
\tilde{\Fpq}=\sqrt{\Fpq^2+\rho_Xa_{PQ}^2} ,
\end{equation}
which is equivalent to rescale the Peccei-Quinn scale:
\begin{equation}
\frac{1}{2}\pmud \tilde{S_a} \pmuu \tilde{S_a} = \frac{\tilde{\Fpq}^2}{2}\pmud N\pmuu N\dgg ,~~~~ N = e^{i\frac{S_a}{\Fpq}} = e^{i\frac{\tilde{S_a}}{\tilde{\Fpq}}} .
\end{equation}
Moreover, the interaction term between $U$ and $X$ in the interpolating model becomes exactly the ``instantonic'' interaction term of the EL$_\sigma$ model with the addition of the axion, i.e.,
\begin{equation}
\frac{\kappa_1}{2\sqrt{2}}[X^\dagger \det U + X\det U^\dagger] \rightarrow
\kappa_I[N^{a_{PQ}}\det U + (N^{\dagger})^{a_{PQ}} \det U^{\dagger}] ,
\end{equation} 
with the identification $\kappa_I\equiv\frac{\kappa_1\sqrt{\rho_X}}{2\sqrt{2}}$.

By virtue of the correspondence relations \eqref{IM to WDV} and \eqref{IM to ELSM}, it is sufficient to make all the calculations within the interpolating model since the results for the WDV and EL$_\sigma$ models are easily obtained by making the above-mentioned proper limits.
In the next section we shall take advantage of this last consideration by computing the axion mass for the interpolating model in the large-$\Fpq$ limit in order to extract the QCD topological susceptibility at finite temperature, both above and below the chiral transition, and then deduce the corresponding results for the WDV and EL$_\sigma$ effective models.

\section{Axion mass and topological susceptibility at finite temperature}
\label{sec3}

In this section we shall compute the axion mass at the leading order in $1/\Fpq$, exploiting the fact that the determinant of the full squared-mass matrix of the model necessarily vanishes in the limit $\Fpq\rightarrow\infty$, since in this limit the axion becomes massless. This means that the axion squared mass (and thus the QCD topological susceptibility) can be obtained, at the leading order in $1/\Fpq$, from the ratio of the determinant of the full squared-mass matrix and the determinant of the squared-mass matrix without the axion (which coincides with the minor with nonzero entries resulting from taking $\Fpq\rightarrow\infty$).

\subsection{Below the chiral transition ($T<T_c$)}

Using the following parametrization for the VEVs of the fields $U$, $X$, and $N$ (being the quark-mass matrix $M$ diagonal, we can take $\langle U \rangle$ to be diagonal too):
\begin{equation}
\langle U_{ij} \rangle = \rho_ie^{i\phi_i}\delta_{ij},
\quad \langle X \rangle = \alpha e^{i\beta},
\quad \langle N \rangle = e^{i\phi},
\end{equation}
and (following the notation of Refs. \cite{EM2019,MM2013}) writing the parameter $\rho_X$ as follows:
\begin{equation}
\rho_X \equiv \frac{F_X^2}{2}>0,
\end{equation}
the potential \eqref{IM+axion_potential} (evaluated on the VEVs of the fields) turns out to be
\begin{equation}
\begin{split}
V = & -\frac{B_m}{\sqrt{2}}\sum_im_i\rho_i\cos\phi_i+\frac{\lambda_{\pi}^2}{4}\sum_i\left(\rho_i^2-\rho_{\pi}\right)^2+\frac{\lambda_{\pi}^{'2}}{4}\Big(\sum_i\rho_i^2\Big)^2+\frac{\lambda_X^2}{4}\left(\alpha^2-\frac{F_X^2}{2}\right)^2\\
&-\frac{\kappa_1\alpha}{\sqrt{2}}\cos\Big(\beta-\sum_i\phi_i\Big)\prod_i\rho_i+\frac{A}{2}\Big(\omega_1\sum_i\phi_i+(1-\omega_1)\beta+a_{PQ}\phi\Big)^2 ,
\end{split}
\end{equation}
from which the stationary-point equations read
\begin{equation}
\label{sysbtint}
\left\{
\begin{aligned}
\frac{\partial V}{\partial \rho_i} = ~ & -\frac{B_m}{\sqrt{2}} m_i \cos\phi_i+\rho_i\Big(\lambda_{\pi}^{'2}\sum_j\rho_j^2+\lambda_{\pi}^2\rho_i^2-\lambda_{\pi}^2\rho_{\pi}\Big)\\
&-\frac{\kappa_1\alpha}{\sqrt{2}}\cos\Big(\beta-\sum_j\phi_j\Big)\prod_{j\neq i}\rho_j=0 ,\\
\frac{\partial V}{\partial \phi_i} = ~ & \frac{B_m}{\sqrt{2}}m_i\rho_i\sin\phi_i-\frac{\kappa_1\alpha}{\sqrt{2}}\sin\Big(\beta-\sum_j\phi_j\Big)\prod_j\rho_j\\
&+\omega_1\Big(\omega_1\sum_j\phi_j+(1-\omega_1)\beta+a_{PQ}\phi\Big)=0 ,\\
\frac{\partial V}{\partial \alpha} = ~ & \lambda_X^2\left(\alpha^2-\frac{F_X^2}{2}\right)\alpha-\frac{\kappa_1}{\sqrt{2}}\cos\Big(\beta-\sum_j\phi_j\Big)\prod_j\rho_j=0 ,\\
\frac{\partial V}{\partial \beta}= ~ & \frac{\kappa_1\alpha}{\sqrt{2}}\sin\Big(\beta-\sum_j\phi_j\Big)\prod_j\rho_j\\
&+(1-\omega_1)A\Big(\omega_1\sum_j\phi_j+(1-\omega_1)\beta+a_{PQ}\phi\Big)=0 ,\\
\frac{\partial V}{\partial \phi} = ~ & a_{PQ}A\Big(\omega_1\sum_j\phi_j+(1-\omega_1)\beta+a_{PQ}\phi\Big)=0 .
\end{aligned}
\right.
\end{equation}
Since now $CP$ is an exact symmetry, the VEVs of the pseudoscalar fields ($\phi_i$, $\beta$, $\phi$) must vanish. In addition, there can be no mixing between scalar and pseudoscalar degrees of freedom, so that we can look at the squared-mass matrix for the (canonically normalized) pseudoscalar fields $\pi_{11}=\rho_1\phi_1,~\pi_{22}=\rho_2\phi_2,\ldots,S_X=\alpha\beta,~S_a=\Fpq\phi$ alone, which turns out to be
\begin{equation}
\mathcal{M}^2 = F H F ,
\end{equation}
where $F = \diag(\rho_1^{-1},\ldots,\rho_{n_l}^{-1},\alpha^{-1},\Fpq^{-1})$ and
\[
H = \begin{pmatrix}
S_1+R+\omega_1^2A & R+\omega_1^2A & \cdots & -R+\omega_1(1-\omega_1)A & \omega_1 a_{PQ}A\\
R+\omega_1^2A & S_2+R+\omega_1^2A & \cdots & -R+\omega_1(1-\omega_1)A & \omega_1 a_{PQ}A\\
\vdots & \vdots & \ddots & \vdots & \vdots \\
-R+\omega_1(1-\omega_1)A & -R+\omega_1(1-\omega_1)A & \cdots & R+(1-\omega_1)^2A & (1-\omega_1)a_{PQ}A \\
\omega_1 a_{PQ}A & \omega_1 a_{PQ}A & \cdots & (1-\omega_1)a_{PQ}A & a_{PQ}^2A
\end{pmatrix} ,
\]
having defined
\begin{equation}
\label{R_and_S_i}
R \equiv \frac{\kappa_1\alpha}{\sqrt{2}}\prod_i\rho_i,~~~~ S_i \equiv \frac{B_m}{\sqrt{2}}m_i\rho_i .
\end{equation}
We find that
\begin{equation}
\det \mathcal{M}^2 = (\det F)^2 a_{PQ}^2AR\prod_iS_i = \frac{\kappa_1 a_{PQ}^2 A}{\sqrt{2} \alpha \Fpq^2} \left(\frac{B_m}{\sqrt{2}}\right)^{n_l}\prod_i m_i ,
\end{equation}
while the determinant of the minor obtained by removing the last row and the last column is
\begin{equation}
\det \mathcal{M}^2_< =
\frac{\kappa_1 A}{\sqrt{2} \alpha} \left(\frac{B_m}{\sqrt{2}}\right)^{n_l}
\left( \frac{1}{A}+\frac{(1-\omega_1)^2}{R}+\sum_j\frac{1}{S_j} \right)
\prod_i m_i ,
\end{equation}
so that, at the leading order in $1/\Fpq$,
\begin{equation}
\label{eq_ax_bel}
m^2_{axion} \simeq \frac{\det \mathcal{M}^2}{\det \mathcal{M}^2_<} = \frac{a_{PQ}^2}{\Fpq^2}\frac{1}{\frac{1}{A}+\frac{(1-\omega_1)^2}{R}+\sum_j\frac{1}{S_j}} .
\end{equation}
By virtue of Eq. \eqref{axmass}, this yields the following expression for the topological susceptibility:
\begin{equation}
\label{eq_chi_bel}
\chi_{QCD} = \frac{1}{\frac{1}{A}+\frac{(1-\omega_1)^2}{R}+\sum_j\frac{1}{S_j}} ,
\end{equation}
where $\rho_i$ and $\alpha$, contained in $R$ and $S_i$, solve the following stationary-point equations:
\begin{equation}
\left\{
\begin{aligned}
\frac{\partial V}{\partial \rho_i} = & -\frac{B_m}{\sqrt{2}}m_i+\rho_i\left(\lambda_{\pi}^{'2}\sum_{j=1}^{n_l}\rho_j^2+\lambda_{\pi}^2\rho_i^2-\lambda_{\pi}^2\rho_{\pi}\right)-\frac{\kappa_1\alpha}{\sqrt{2}} \prod_{j\neq i}\rho_j=0 ,\\
\frac{\partial V}{\partial \alpha} = & \lambda_X^2\left(\alpha^2-\frac{F_X^2}{2}\right)\alpha-\frac{\kappa_1}{\sqrt{2}}\prod_{j=1}^{n_l}\rho_j=0 .
\end{aligned}
\right.
\end{equation}
Finally, making use of the relations \eqref{IM to WDV} and \eqref{IM to ELSM} found at the end of Sec. \ref{sec2}, we can immediately write down the expressions which one obtains for the topological susceptibility using the WDV and the EL$_\sigma$ effective models for $T<T_c$.
\begin{itemize}
\item WDV model:
\begin{equation}
\label{eq_chi_WDV_bel}
\chi_{QCD}^{(WDV)} = \frac{1}{\frac{1}{A}+\sum_j\frac{1}{S_j}}
= \frac{A}{1+A\sum_j\frac{\sqrt{2}}{B_m m_j\rho_j}} ,
\end{equation}
\noindent where the parameters $\rho_i$ solve the equations
\begin{equation}
-\frac{B_m}{\sqrt{2}} m_i+\rho_i\left(\lambda_{\pi}^{'2}\sum_{j=1}^{n_l}\rho_j^2+\lambda_{\pi}^2\rho_i^2-\lambda_{\pi}^2\rho_{\pi}\right)=0 .
\end{equation}
\item EL$_\sigma$ model:
\begin{equation}
\label{eq_chi_ELSM_bel}
\chi_{QCD}^{(EL_\sigma)} = \frac{1}{\frac{1}{R}+\sum_j\frac{1}{S_j}}
= \frac{2\kappa_I\prod_i\rho_i}{1+2\kappa_I\prod_i\rho_i\sum_j\frac{\sqrt{2}}{B_m m_j\rho_j}} ,
\end{equation}
\noindent where $\kappa_I \equiv \frac{\kappa_1\alpha}{2\sqrt{2}} = \frac{\kappa_1 F_X}{4}$ and the parameters $\rho_i$ solve the equations
\begin{equation}
-\frac{B_m}{\sqrt{2}} m_i+\rho_i\left(\lambda_{\pi}^{'2}\sum_{j=1}^{n_l}\rho_j^2+\lambda_{\pi}^2\rho_i^2-\lambda_{\pi}^2\rho_{\pi}\right)-2\kappa_I \prod_{j\neq i}\rho_j=0 .
\end{equation}
\end{itemize}
The results \eqref{eq_chi_bel}, \eqref{eq_chi_WDV_bel}, and \eqref{eq_chi_ELSM_bel} generalize the corresponding results found in Ref. \cite{LM2018}, studying the $\theta$ dependence of the vacuum energy density (free energy) and using the \emph{nonlinear} versions of the various effective models, in which the scalar degrees of freedom had been simply integrated out, by taking the limit $\lambda_\pi^2 \to \infty$ (\emph{decoupling limit}): in this limit the solutions of the stationary-point equations simply reduce to $\rho_i = \sqrt{\rho_\pi} \equiv \frac{F_\pi}{\sqrt{2}}$, i.e., $\langle U \rangle = \frac{F_\pi}{\sqrt{2}}\mathbf{I}$, where $F_\pi$ is the so-called \emph{pion decay constant}.\\
In the Appendix, we shall give a numerical evaluation of the expressions \eqref{eq_chi_WDV_bel} and \eqref{eq_chi_ELSM_bel} for the topological susceptibility at zero temperature, using the WDV and EL$_\sigma$ models in the physical case $n_l=3$: the results will be compared with the corresponding results found in Ref. \cite{LM2018} (using the nonlinear effective models) and with other estimates present in the literature.

\subsection{Above the chiral transition ($T>T_c$)}

As already recalled in the previous section, in order to avoid a singular behavior of the anomalous term in the potential \eqref{IM_potential} above the chiral-transition temperature $T_c$, where the VEV of the mesonic field $U$ vanishes (in the chiral limit $M=0$), we must assume that \cite{EM1994,MM2013} $\omega_1 (T\geq T_c)=0$. In this regime of temperatures, therefore, the potential is given by
\begin{equation}
\begin{split}
V(U,U^{\dagger},X, & X^{\dagger}, N, N^{\dagger})=~ -\frac{B_m}{2\sqrt{2}} \Tr [M(U+ U^{\dagger})] +\frac{\lambda _{\pi}^2}{4} \Tr [(U^{\dagger}U-\rho_{\pi}\mathbf{I})^2]\\
& +\frac{\lambda _{\pi}^{'2}}{4} (\Tr [U^{\dagger}U])^2 +\frac{\lambda_X^2}{4}[XX^{\dagger}-\rho_X]^2-\frac{\kappa_1}{2\sqrt{2}}(X \det U^{\dagger}+ X^{\dagger}\det U)\\
& + \frac{A}{2}\left(\frac{i}{2}[\log X- \log X^{\dagger} ] +\frac{ia_{PQ}}{2}[\log N- \log N^{\dagger}]\right)^2 .
\end{split}
\end{equation}
Using, now, the following parametrization for the VEVs of the fields $U$, $X$, and $N$:
\begin{equation}
\langle U_{ij} \rangle = (\rho_i+i\eta_i)\delta_{ij},
\quad \langle X \rangle = \alpha e^{i\beta},
\quad \langle N \rangle = e^{i\phi},
\end{equation}
and (following, as usual, the notation of Refs. \cite{EM2019,MM2013}) writing the parameters $\rho_\pi$ and $\rho_X$ for $T>T_c$ as follows:
\begin{equation}
\rho_{\pi} \equiv -\frac{B_{\pi}^2}{2}<0,\quad \rho_X \equiv \frac{F_X^2}{2}>0,
\end{equation}
the potential (evaluated on the VEVs of the fields) turns out to be
\begin{equation}
\begin{split}
V = & -\frac{B_m}{\sqrt{2}}\sum_{i=1}^{n_l}m_i\rho_i+\frac{\lambda_{\pi}^2}{4}\sum_{i=1}^{n_l}(\rho_i^2+\eta_i^2)^2+\frac{\lambda_{\pi}^2B_{\pi}^2}{4}\sum_{i=1}^{n_l}(\rho_i^2+\eta_i^2)+\frac{\lambda_{\pi}^{'2}}{4}\left(\sum_{i=1}^{n_l}(\rho_i^2+\eta_i^2)\right)^2\\
&+\frac{\lambda_X^2}{4}\left(\alpha^2-\frac{F_X^2}{2}\right)^2-\frac{\kappa_1\alpha}{2\sqrt{2}}\left(e^{i\beta}\prod_{i=1}^{n_l}(\rho_i-i\eta_i)+e^{-i\beta}\prod_{i=1}^{n_l}(\rho_i+i\eta_i)\right)\\
&+\frac{A}{2}\left(\beta+a_{PQ}\phi\right)^2 .
\end{split}
\end{equation}
Once more, the inclusion of the axion implies $CP$ conservation and, as a consequence, the vanishing of all the VEVs of the pseudoscalar degrees of freedom ($\eta_i$, $\beta$, $\phi$) and of their mixings with the scalar degrees of freedom, whose VEVs can be obtained from the corresponding stationary-point equations:
\begin{equation}
\label{last_sys}
\left\{
\begin{aligned}
\frac{\partial V}{\partial \rho_i} = & -\frac{B_m}{\sqrt{2}}m_i+\rho_i\left(\lambda_{\pi}^{'2}\sum_{j=1}^{n_l}\rho_j^2+\lambda_{\pi}^2\rho_i^2+\frac{\lambda_{\pi}^2 B_{\pi}^2}{2}\right)-\frac{\kappa_1\alpha}{\sqrt{2}} \prod_{j\neq i}\rho_j=0 ,\\
\frac{\partial V}{\partial \alpha} = & ~ \lambda_X^2\left(\alpha^2-\frac{F_X^2}{2}\right)\alpha-\frac{\kappa_1}{\sqrt{2}}\prod_{j=1}^{n_l}\rho_j=0 .
\end{aligned}
\right.
\end{equation}
The computation of the second derivatives of the potential at the minimum of $V$ leads to the following squared-mass matrix for the pseudoscalar field $\eta_1,~\eta_2,\ldots,S_X=\alpha\beta,~S_a=\Fpq\phi$:
\begin{equation}
\mathcal{M}^2 = 
\left(
\begin{array}{cccccc}
K+\lambda_{\pi}^2\rho_1^2 & \frac{\kappa_1\alpha}{\sqrt{2}}\prod_{k\neq 1, 2}\rho_k & \cdots & -\frac{\kappa_1}{\sqrt{2}}\prod_{k\neq 1}\rho_k & 0 \\
\frac{\kappa_1\alpha}{\sqrt{2}}\prod_{k\neq 1, 2}\rho_k & K +\lambda_{\pi}^2\rho_2^2 & \cdots & -\frac{\kappa_1}{\sqrt{2}}\prod_{k\neq 2}\rho_k & 0 \\
 \vdots & \vdots & \ddots & \vdots & \vdots \\
-\frac{\kappa_1}{\sqrt{2}}\prod_{k\neq 1}\rho_k & -\frac{\kappa_1}{\sqrt{2}}\prod_{k\neq 2}\rho_k & \cdots & \frac{\kappa_1}{\sqrt{2}\alpha}\prod_i\rho_i + \frac{A}{\alpha^2} & \frac{a_{PQ} A}{\alpha \Fpq} \\
0 & 0 & \cdots & \frac{a_{PQ} A}{\alpha \Fpq} & \frac{a_{PQ}^2 A}{\Fpq^2} \\
\end{array}
\right) ,
\end{equation}
where $K \equiv \frac{\lambda_{\pi}^2 B_{\pi}^2}{2}+\lambda_{\pi}^{'2}\sum_j\rho_j^2$. Its determinant is given by
\begin{equation}
\det \mathcal{M}^2 = \frac{A\kappa_1}{\sqrt{2}\alpha}\frac{a_{PQ}^2}{\Fpq^2} \prod_i\left[\rho_i\left(K+\lambda_{\pi}^2\rho_i^2\right)-\frac{\kappa_1\alpha}{\sqrt{2}}\prod_{j\neq i}\rho_j\right]
= \frac{\kappa_1 a_{PQ}^2 A}{\sqrt{2} \alpha \Fpq^2} \left(\frac{B_m}{\sqrt{2}}\right)^{n_l}\prod_i m_i ,
\end{equation}
where Eq. (\ref{last_sys}) has been used.
Instead, the determinant of the minor obtained by removing the last row and the last column (which survives the $\Fpq \rightarrow \infty$ limit) is found to be [making again use of Eq. \eqref{last_sys}]
\begin{equation}
\det \mathcal{M}^2_< =
\frac{\kappa_1 A}{\sqrt{2} \alpha} \left(\frac{B_m}{\sqrt{2}}\right)^{n_l}
\left( \frac{1}{A}+\frac{1}{R}+\sum_j\frac{1}{S_j} \right) \prod_i m_i ,
\end{equation}
where $R$ and $S_i$ are defined as in the previous subsection; see Eq. \eqref{R_and_S_i}.
Therefore, the axion mass is given by (at the leading order in $1/\Fpq$)
\begin{equation}
\label{eq_ax_ab}
m^2_{axion} \simeq \frac{\det \mathcal{M}^2}{\det \mathcal{M}^2_<} = \frac{a_{PQ}^2}{\Fpq^2}\frac{1}{\frac{1}{A}+\frac{1}{R}+\sum_j\frac{1}{S_j}} .
\end{equation}
From this, by virtue of Eq. \eqref{axmass}, we derive the following expression for the QCD topological susceptibility above the chiral transition:
\begin{equation}
\label{eq_chi_ab}
\chi_{QCD} = \frac{1}{\frac{1}{A}+\frac{1}{R}+\sum_j\frac{1}{S_j}} ,
\end{equation}
which is formally identical to the expression \eqref{eq_chi_bel} with $\omega_1=0$, but with the difference that now $\rho_i$ and $\alpha$ must solve the stationary-point equations \eqref{last_sys}.

Finally, making use of the relation \eqref{IM to ELSM} found at the end of Sec. \ref{sec2}, we can immediately write down the expression which one obtains for the topological susceptibility using the EL$_\sigma$ effective model for $T>T_c$:
\begin{equation}
\label{eq_chi_ELSM_ab}
\chi_{QCD}^{(EL_\sigma)} = \frac{1}{\frac{1}{R}+\sum_j\frac{1}{S_j}}
= \frac{2\kappa_I\prod_i\rho_i}{1+2\kappa_I\prod_i\rho_i\sum_j\frac{\sqrt{2}}{B_m m_j\rho_j}} ,
\end{equation}
\noindent where $\kappa_I \equiv \frac{\kappa_1\alpha}{2\sqrt{2}} = \frac{\kappa_1 F_X}{4}$ and the parameters $\rho_i$ solve the equations
\begin{equation}
\label{last_sys_ELSM}
-\frac{B_m}{\sqrt{2}}m_i+\rho_i\left(\lambda_{\pi}^{'2}\sum_{j=1}^{n_l}\rho_j^2+\lambda_{\pi}^2\rho_i^2+\frac{\lambda_{\pi}^2 B_{\pi}^2}{2}\right)-2\kappa_I \prod_{j\neq i}\rho_j=0 .
\end{equation}
The results \eqref{eq_chi_ab} and \eqref{eq_chi_ELSM_ab} generalize the corresponding results which were derived in Ref. \cite{EM2019}, studying the $\theta$ dependence of the vacuum energy density (free energy) at the first nontrivial order in an expansion in the quark masses.
Solving the stationary-point equations \eqref{last_sys} and \eqref{last_sys_ELSM} at the leading order in the quark masses, one finds that, in the case $n_l=3$,
$\rho_i \simeq \frac{\sqrt{2} B_m}{\lambda_\pi^2 B_\pi^2} m_i$ and $\alpha \simeq \frac{F_X}{\sqrt{2}}$, so that, substituting in Eqs. \eqref{eq_chi_ab} and \eqref{eq_chi_ELSM_ab}, one finds the same approximate expression already derived in Ref. \cite{EM2019} for the topological susceptibility:
\begin{equation}
\chi \simeq \frac{\kappa_1 F_X}{2}
\left( \frac{\sqrt{2} B_m}{\lambda_\pi^2 B_\pi^2} \right)^{n_l} \det M
= 2\kappa_I
\left( \frac{\sqrt{2} B_m}{\lambda_\pi^2 B_\pi^2} \right)^{n_l} \det M .
\end{equation}
A similar result occurs also in the special case $n_l=2$. In this case, solving the stationary-point equations \eqref{last_sys} and \eqref{last_sys_ELSM} at the leading order in the quark masses, one finds that
$\rho_1 \simeq \sqrt{2} B_m \frac{\lambda_\pi^2 B_\pi^2 m_1 + \kappa_1 F_X m_2}{\lambda_\pi^4 B_\pi^4 - \kappa_1^2 F_X^2}$,
$\rho_2 \simeq \sqrt{2} B_m \frac{\lambda_\pi^2 B_\pi^2 m_2 + \kappa_1 F_X m_1}{\lambda_\pi^4 B_\pi^4 - \kappa_1^2 F_X^2}$,
and $\alpha \simeq \frac{F_X}{\sqrt{2}}$, so that, substituting in Eqs. \eqref{eq_chi_ab} and \eqref{eq_chi_ELSM_ab}, one finds also in this case the same approximate expression already derived in Ref. \cite{EM2019} for the topological susceptibility:
\begin{equation}
\chi \simeq
\frac{\kappa_1 F_X B_m^2}{\lambda_\pi^4 B_\pi^4 - \kappa_1^2 F_X^2} m_u m_d
= \frac{4 \kappa_I B_m^2}{\lambda_\pi^4 B_\pi^4 - 16\kappa_I^2} m_u m_d .
\end{equation}
Further comments on the ``exact'' expressions \eqref{eq_chi_ab} and \eqref{eq_chi_ELSM_ab}, derived in this paper for the topological susceptibility for $T>T_c$, will be made in the next section.

\section{Summary of the results and conclusions}
\label{sec4}

In this paper we have computed the axion mass and, from this (exploiting the well-known formula \eqref{axmass}, valid in the limit of very large $\Fpq$, i.e., $\Fpq \gg \Lambda_{QCD}$), we have derived an expression for the QCD topological susceptibility in the finite-temperature case, both below and above the chiral phase transition at $T_c$, making use of a chiral effective Lagrangian model, the so-called interpolating model, which includes the axion, the scalar and pseudoscalar mesons and implements the $U(1)$ axial anomaly of the fundamental theory.
The choice of this model (described in detail in Sec. \ref{sec2}) is due to its ``regularity'' around the chiral phase transition (i.e., it is well defined also above $T_c$) and to the fact that the other known chiral effective Lagrangian models, namely the WDV and EL$_\sigma$ models (and the corresponding results for the axion mass and the QCD topological susceptibility, both below and above $T_c$), can be obtained (as is shown at the end of Sec. \ref{sec2}) by taking proper formal limits of the interpolating model (and its results).

As we can see by giving a closer look at the results obtained for the topological susceptibility in the previous section, the expressions \eqref{eq_chi_bel} (for $T<T_c$) and \eqref{eq_chi_ab} (for $T>T_c$) are formally the same since they are both given by
\begin{equation}
\label{chi}
\left\{
\begin{aligned}
& \chi_{QCD} = \frac{1}{\frac{1}{A}+\frac{(1-\omega_1)^2}{R}+\sum_j\frac{1}{S_j}} ,\\
& R \equiv \frac{\kappa_1\alpha}{\sqrt{2}}\prod_i\rho_i, \quad S_i \equiv \frac{B_m}{\sqrt{2}}m_i\rho_i,
\end{aligned}
\right.
\end{equation}
where the VEVs $\rho_i$ and $\alpha$ are obtained by solving the following stationary-point equations:
\begin{equation}
\left\{
\begin{aligned}
&-\frac{B_m}{\sqrt{2}} m_i+\rho_i\left(\lambda_{\pi}^{'2}\sum_{j=1}^{n_l}\rho_j^2+\lambda_{\pi}^2\rho_i^2-\lambda_{\pi}^2\rho_{\pi}\right)-\frac{\kappa_1\alpha}{\sqrt{2}}\prod_{j\neq i}\rho_j=0 ,\\
&\lambda_X^2\left(\alpha^2-\frac{F_X^2}{2}\right)\alpha-\frac{\kappa_1}{\sqrt{2}}\prod_{i=1}^{n_l}\rho_i=0 ,
\end{aligned}
\right.
\end{equation}
with the following temperature dependence of the parameters $\rho_\pi$ and $\omega_1$:
\begin{equation}
\rho_\pi(T<T_c) > 0,\quad \rho_\pi(T>T_c) \equiv -\frac{B_{\pi}^2}{2} < 0, \quad \omega_1(T>T_c)=0 .
\end{equation}
From this result, making use of the relations \eqref{IM to WDV} and \eqref{IM to ELSM} found at the end of Sec. \ref{sec2}, we immediately derive the expressions for the topological susceptibility using the WDV and the EL$_\sigma$ effective models for $T<T_c$, as well as the expression for the topological susceptibility using the EL$_\sigma$ effective model for $T>T_c$.

Concerning the results for $T<T_c$, the expressions \eqref{eq_chi_bel}, \eqref{eq_chi_WDV_bel}, and \eqref{eq_chi_ELSM_bel} generalize the corresponding results found in Ref. \cite{LM2018}, studying the $\theta$ dependence of the vacuum energy density (free energy) and using the nonlinear versions of the various effective models, in which the scalar degrees of freedom had been simply integrated out, by taking the decoupling limit $\lambda_\pi^2 \to \infty$.\\
In the Appendix [Eqs. \eqref{chi_WDV_T=0_result} and \eqref{chi_ELSM_T=0_result}], we have given a numerical evaluation of the expressions \eqref{eq_chi_WDV_bel} and \eqref{eq_chi_ELSM_bel} for the topological susceptibility at $T=0$, using the WDV and EL$_\sigma$ models in the physical case $n_l=3$.
The results,
\begin{equation}\label{chi_T=0_results}
\chi^{(WDV)}_{QCD} = (75.7 \pm 0.2 ~\text{MeV})^4,~~~~
\chi^{(EL_\sigma)}_{QCD} = (75.8 \pm 0.2 ~\text{MeV})^4,
\end{equation}
have been compared with the corresponding results found in Ref. \cite{LM2018} using the nonlinear effective models: the inclusion of the scalar degrees of freedom leads to a non-negligible difference between the above-reported results and those obtained in Ref. \cite{LM2018} in the decoupling limit (i.e., simply integrating out the scalar degrees of freedom).
Moreover, the two above-reported results are perfectly consistent with each other and in agreement with the available most accurate lattice determination of $\chi_{QCD}$ and also with the results obtained using the chiral perturbation theory with $n_l=2$ light flavors up to the next-to-next-to-leading order (see the Appendix).

Concerning the results for $T>T_c$, the expressions \eqref{eq_chi_ab} and \eqref{eq_chi_ELSM_ab} generalize the corresponding results which were derived in Ref. \cite{EM2019}, studying the $\theta$ dependence of the vacuum energy density (free energy) at the first nontrivial order in an expansion in the quark masses. Even if, of course, in this case we cannot make any more quantitative statements (like we have done, instead, in the case at $T=0$), nevertheless, we want to make some remarks concerning the question of the temperature and quark-mass dependence.\\
If we assume (as it appears reasonable on the basis of our knowledge on the role of instantons at finite temperature) that the $U(1)$ axial condensate vanishes at high temperatures with a certain power law in $T$, i.e., $\alpha$ (or, better, $\kappa_1 \alpha$) $\sim T^{-k}$ (for some positive coefficient $k$), we would be tempted to conclude from Eq. \eqref{eq_chi_ab} that also $\chi_{QCD}$ vanishes at high temperatures in the same way, i.e.,\footnote{We must also assume that the other quantities $S_i$ have a much milder dependence on $T$, and, moreover, that $R \ll A$, which is equivalent to $\chi_{QCD} \ll A$. (Since also $A$ is expected to vanish at large temperatures, this means that $A \sim T^{-k_A}$, with $k_A \le k$: in the opposite case $k_A > k$, we would obtain that $\chi_{QCD} \simeq A \sim T^{-k_A}$.) At least at $T=0$, this condition is reasonably satisfied, since in that case one identifies $A$ with the \emph{pure-gauge} topological susceptibility and (see the Appendix) $\chi(T=0) \simeq (75~{\rm MeV})^4$, $A(T=0) \simeq (180~{\rm MeV})^4$.
However, at finite temperature, it is not even clear if, in our phenomenological Lagrangian for the interpolating model, the parameter $A(T)$ can be simply identified with the pure-gauge topological susceptibility.}
\begin{equation}
\chi_{QCD} \mathop{\simeq}^{?} R \sim T^{-k} ,
\end{equation}
being (at the leading order in the quark masses
$\rho_i \simeq \frac{\sqrt{2} B_m}{\lambda_\pi^2 B_\pi^2} m_i$ and $\alpha \simeq \frac{F_X}{\sqrt{2}}$;
it is reasonable to assume that this approximation makes sense for $T - T_c \gg m_f$, but \emph{not} for $T$ very close to $T_c$, i.e., for $T - T_c \lesssim m_f$)\footnote{Thanks to the vanishing of the $U(1)$ axial condensate $\alpha \simeq \frac{F_X}{\sqrt{2}}$, it is easy to see that this result applies for any $n_l$, including the special case $n_l=2$.}
\begin{equation}
R \equiv \frac{\kappa_1\alpha}{\sqrt{2}}\prod_i\rho_i \simeq \frac{\kappa_1 F_X}{2} \left(\frac{\sqrt{2} B_m}{\lambda_\pi^2 B_\pi^2}\right)^{n_l} \prod_i m_i .
\end{equation}
We observe that, of course, this same result would be obtained also in the case of the EL$_\sigma$ model, starting from Eq. \eqref{eq_chi_ELSM_ab}, with the usual identification $\kappa_I \equiv \frac{\kappa_1\alpha}{2\sqrt{2}} = \frac{\kappa_1 F_X}{4}$.
In this way, both the temperature dependence of $\chi_{QCD}$ and its quark-mass dependence (proportional to $\det M$) would turn out to be in agreement with the results found using the so-called dilute instanton-gas approximation (DIGA) \cite{GPY1981}, with $k = \frac{11}{3}N_c + \frac{1}{3}n_l - 4 = 7 + \frac{1}{3}n_l$.
The problem with the above-reported argumentations is, of course, that the use of an effective model in terms of mesonic excitations, while being probably still legitimate immediately above $T_c$, is surely no longer valid for very high temperatures ($T \gg T_c$), where the quark and gluon degrees of freedom of the fundamental theory become more and more relevant.
In other words, it is not obvious at all that the range of validity (in temperature) of Eq. \eqref{eq_chi_ab} has an overlap with the range of validity of the DIGA prediction. (For example, in Ref. \cite{FH2016}, investigating the quantum and thermal fluctuations in the EL$_\sigma$ model and their effect on the chiral anomaly, it was found that mesonic fluctuations cause an increase, rather than a decrease, of the parameter $\kappa_I$ for temperatures $T$ toward $T_c$, and the authors conclude that it remains an important question whether the temperature dependence of $\kappa_I$ that arises from instanton effects can compete with mesonic fluctuations.)

In this respect, recent lattice investigations have shown contrasting results.
Some first studies \cite{lattice_1,lattice_2} have found appreciable deviations from the DIGA prediction for temperatures $T$ up to about 600 MeV, while later studies \cite{lattice_3,lattice_4,lattice_5,lattice_6,lattice_7} have shown a substantial agreement with the DIGA prediction, in a range of temperatures which in some cases starts right above $T_c$, in other cases starts from 2 or 3 times $T_c$ and goes up to a few GeVs.
The situation is thus not yet fully settled and calls for further and more accurate studies (in this respect, see also Ref. \cite{DDSX2017}).
Moreover, as far as we know, the question of the quark-mass dependence of $\chi_{QCD}$ at high temperatures (above $T_c$) has not yet been investigated on the lattice.

Therefore, future works (both analytical and numerical) will be necessary to shed more light on these questions.
We also recall that, by virtue of the relation \eqref{axmass}, a more accurate knowledge of $\chi_{QCD}(T)$ in the high-temperature regime (at the GeV scale or above) would allow one to obtain a more precise estimate of the coupling constant $\Fpq$ (or, better, $\Fpq/a_{PQ}$), assuming that the axion is the main component of dark matter (through the so-called ``misalignment mechanism'' \cite{PWW1983,AS1983,DF1983}): this in turn would allow one to obtain a more precise estimate of the axion mass today (at $T=0$), a useful (if not necessary) input for all present and future experimental searches for the axion.


\newpage

\renewcommand{\thesection}{}
\renewcommand{\thesubsection}{A.\arabic{subsection} }
 
\pagebreak[3]
\setcounter{section}{1}
\setcounter{equation}{0}
\setcounter{subsection}{0}
\setcounter{footnote}{0}

\begin{flushleft}
{\Large\bf \thesection Appendix: Numerical results for the topological susceptibility at $T=0$}
\end{flushleft}

\renewcommand{\thesection}{A}

\noindent
In this Appendix, we shall give a numerical evaluation of the expressions \eqref{eq_chi_WDV_bel} and \eqref{eq_chi_ELSM_bel} for the topological susceptibility at zero temperature, using the WDV and EL$_\sigma$ models in the physical case $n_l=3$.
In order to do this, we need to know the values of the various parameters which appear in these expressions: $\rho_i$, $B_m m_i$, $A$, and $\kappa_I$.\\
We first consider the parameters $\rho_i$'s, which appear in the vacuum expectation value $\langle U\rangle = \diag(\rho_1,\rho_2,\rho_3)$.
Using for $U$ the following linear parametrization:
\begin{equation}
U = \sqrt{2}(\sigma_a+i\pi_a)T_a ,
\end{equation}
where $T_a$ ($a=0,\ldots, n_l^2-1$; $T_0=\frac{1}{\sqrt{2n_l}}\mathbf{I}$)
are the usual $U(n_l) = U(1) \otimes SU(n_l)$ generators, with the normalization $\Tr[T_a T_b] = \frac{1}{2} \delta_{ab}$, we can write the vacuum expectation value of $U$ as: $\langle U \rangle = \sqrt{2} (\langle\sigma_0\rangle T_0 + \langle\sigma_3\rangle T_3 + \langle\sigma_8\rangle T_8)$.
It was shown in Ref. \cite{ELSMfiniteT_1a} that, neglecting for simplicity small violations of isospin $SU(2)_V$ (the charged and the neutral pions being almost degenerate in mass), i.e., taking $\langle\sigma_3\rangle \simeq 0$ (that is, neglecting $\langle\sigma_3\rangle$ with respect to $\langle\sigma_0\rangle$ and $\langle\sigma_8\rangle$), the values of the condensates $\langle\sigma_0\rangle$ and $\langle\sigma_8\rangle$ are related, by means of the partially-conserved-axial-vector-current relations, to the values of the pion and kaon decay constants $F_\pi$ and $F_K$:
\begin{equation}
\left\{
\begin{aligned}
\langle\sigma_0\rangle =&~ \frac{F_\pi + 2F_K}{\sqrt{6}} ,\\
\langle\sigma_8\rangle =&~ \frac{2}{\sqrt{3}}(F_\pi - F_K) .
\end{aligned}
\right.
\end{equation}
From $\langle U\rangle = \diag(\rho_1,\rho_2, \rho_3) = \sqrt{2} (\langle\sigma_0\rangle T_0 + \langle\sigma_8\rangle T_8)$, we finally find that
\begin{equation}
\left\{
\begin{aligned}
\rho_1 =&~\rho_2 \equiv \rho = \sqrt{\frac{1}{3}}\langle\sigma_0\rangle+\sqrt{\frac{1}{6}}\langle\sigma_8\rangle = \frac{F_{\pi}}{\sqrt{2}} ,\\
\rho_3 =&~ \sqrt{\frac{1}{3}}\langle\sigma_0\rangle-\sqrt{\frac{2}{3}}\langle\sigma_8\rangle = \frac{2F_K-F_{\pi}}{\sqrt{2}} .
\end{aligned}
\right.
\end{equation}
Always in Ref. \cite{ELSMfiniteT_1a} it was shown that the explicit symmetry-breaking term $H \equiv \frac{B_m}{2} M = h_a T_a = h_0 T_0 + h_3 T_3 + h_8 T_8$, for $M = \diag(m_u,m_d,m_s)$,\footnote{In Ref. \cite{ELSMfiniteT_1a} the field $\Phi = \frac{1}{\sqrt{2}}U = (\sigma_a + i\pi_a)T_a$ is used, in place of $U$, with kinetic term $\Tr[\partial_\mu\Phi\partial^\mu\Phi^\dagger] = \frac{1}{2}\Tr[\partial_\mu U \partial^\mu U^\dagger]$ and with an explicit symmetry-breaking term $\Tr[H(\Phi+\Phi^\dagger)] = \frac{B_m}{2\sqrt{2}}\Tr[M(U+U^\dagger)]$, for $H = \frac{B_m}{2}M$.} can be determined in terms of the pion and kaon masses and decay constants through the following relations (always neglecting small $SU(2)_V$ isospin violations, i.e., taking $h_3 = \frac{B_m}{2}(m_u-m_d) \simeq 0$, that is, neglecting $h_3$ with respect to $h_0$ and $h_8$):
\begin{equation}
\left\{
\begin{aligned}
h_0 =&~ \frac{B_m}{\sqrt{6}}(2\tilde{m}+m_s) = \frac{1}{\sqrt{6}}(m_\pi^2 F_\pi + 2m_K^2 F_K) ,\\
h_8 =&~ \frac{B_m}{\sqrt{3}}(\tilde{m}-m_s) = \frac{2}{\sqrt{3}}(m_\pi^2 F_\pi - m_K^2 F_K) ,
\end{aligned}
\right.
\end{equation}
where $\tilde{m} \equiv \frac{m_u+m_d}{2}$. These relations can be inverted to give $B_m \tilde{m} = m_\pi^2 F_\pi$ and $B_m m_s = 2m_K^2 F_K - m_\pi^2 F_\pi$, or, equivalently,
\begin{equation}
m^2_\pi = \frac{B_m}{F_\pi} \tilde{m} ,~~~~
m^2_K = \frac{B_m}{2F_K} (\tilde{m}+m_s) .
\end{equation}
We can obtain more precise relations (to be finally compared with the experimental values of the pion and kaon masses) by adding also an electromagnetic contribution $\Delta m^2_{e.m.}$ to the squared masses of the charged pions and kaons and, moreover, taking into account the up-down mass splitting in the squared masses of the charged and neutral kaons, i.e.,
\begin{equation}
\left\{
\begin{aligned}
m^2_{\pi^{\pm}} =&~ \frac{B_m}{2F_\pi}(m_u+m_d)+\Delta m^2_{e.m.} ,\\
m^2_{\pi^0} =&~ \frac{B_m}{2F_\pi}(m_u+m_d) ,\\
m^2_{K^{\pm}} =&~ \frac{B_m}{2F_K}(m_u+m_s)+\Delta m^2_{e.m.} ,\\
m^2_{K^0,\bar{K}^0} =&~ \frac{B_m}{2F_K}(m_d+m_s) ,
\end{aligned}
\right.
\end{equation}
which can be easily inverted to give $\Delta m^2_{e.m.} = m^2_{\pi^\pm} - m^2_{\pi^0}$ and
\begin{equation}
\left\{
\begin{aligned}
B_m m_u =&~ F_\pi m^2_{\pi^0} - F_K (\Delta m^2_K + \Delta m^2_\pi) ,\\
B_m m_d =&~ F_\pi m^2_{\pi^0} + F_K (\Delta m^2_K + \Delta m^2_\pi) ,\\
B_m m_s =&~ 2F_K m^2_{K^0,\bar{K}^0} - F_\pi m^2_{\pi^0} - F_K (\Delta m^2_K + \Delta m^2_\pi) ,
\end{aligned}
\right.
\end{equation}
where $\Delta m^2_\pi \equiv m^2_{\pi^\pm} - m^2_{\pi_0} (= \Delta m^2_{e.m.})$ and $\Delta m^2_K \equiv m^2_{K^0,\bar{K}^0} - m^2_{K^\pm}$.\footnote{We easily see that in the limit in which $F_K = F_\pi$, i.e., $\rho = \rho_3 = \frac{F_{\pi}}{\sqrt{2}}$, we recover the well-known relations of the leading-order chiral perturbation theory.}\\
For our numerical computations, the following values of the known parameters have been used:
\begin{itemize}
\item $F_{\pi} = 92.1 \pm 1.2$ MeV and $F_K = 110.02\pm 0.28$ MeV for the pion and kaon decay constants (corresponding to the theoretical values reported in Eq. (84.16) in Ref. \cite{PDG} for $f_\pi \equiv \sqrt{2}F_\pi$ and $f_K \equiv \sqrt{2}F_K$), and the known values of the pion and kaon masses (see Ref. \cite{PDG}):
\begin{equation}
\begin{split}
m_{\pi^{\pm}}&~= 139.57061 \pm 0.00024 ~ \text{MeV},\\
m_{\pi^0} &~=134.9770\pm 0.0005 ~ \text{MeV},\\
m_{K^{\pm}}&~=493.677\pm 0.016 ~ \text{MeV},\\
m_{K^0,\bar{K^0}} &~=497.611\pm 0.013 ~ \text{MeV}.
\end{split}
\end{equation} 
\item The parameter $A$ in the interpolating and WDV model is identified (at $T=0$) with the \emph{pure-gauge} topological susceptibility, which has been computed on the lattice: $A=(180\pm 5~ \text{MeV})^4$ (see Ref. \cite{VP2009} and references therein).
\item The parameter $\kappa_I$ in the EL$_\sigma$ model with $n_l=3$ has been computed in Ref. \cite{ELSMfiniteT_1a}: the result, updated with the current values of the experimental inputs, is $\kappa_I = 1721 \pm 50$ MeV.
\end{itemize}
Putting everything together, we obtain the following numerical results for the topological susceptibility $\chi_{QCD}$ at $T=0$ using the WDV and EL$_\sigma$ models in the case $n_l=3$:\footnote{When including the flavor singlet in the effective Lagrangian at $T=0$, we must consider $n_l=3$ for a correct description of the physical world, since the contribution of $\frac{B_m}{\sqrt{2}\rho_3} m_s$ is comparable to $\frac{A}{\rho_3^2} \sim \frac{2A}{F_{\pi}^2}$ in the pseudoscalar squared-mass matrix.}
\begin{itemize}
\item WDV model [see Eq. \eqref{eq_chi_WDV_bel}]:
\begin{equation}\label{chi_WDV_T=0_result}
\chi^{(WDV)}_{QCD} = (75.7 \pm 0.2 ~\text{MeV})^4 ;
\end{equation} 
\item EL$_\sigma$ model [see Eq. \eqref{eq_chi_ELSM_bel}]:
\begin{equation}\label{chi_ELSM_T=0_result}
\chi^{(EL_\sigma)}_{QCD} = (75.8 \pm 0.2 ~\text{MeV})^4 .
\end{equation} 
\end{itemize}
These results should be compared with the corresponding results found in Ref. \cite{LM2018} using the \emph{nonlinear} effective models: $\chi^{(WDV)}_{QCD} = (76.283 \pm 0.106 ~\text{MeV})^4$ and $\chi^{(ENL_\sigma)}_{QCD} = (76.271 \pm 0.085 ~\text{MeV})^4$.
[We also recall here the recent determination obtained in Ref. \cite{Guo2020} using the $SU(3)$ chiral perturbation theory up to the next-to-leading order: $\chi^{(NLO \chi PT_{(3)})}_{QCD} = (76.7 \pm 0.6~ \text{MeV})^4$.]
The inclusion of the scalar degrees of freedom (and, in particular, of the finite splitting $F_K-F_\pi$) leads to a non-negligible difference between the results \eqref{chi_WDV_T=0_result}--\eqref{chi_ELSM_T=0_result} and those obtained in Ref. \cite{LM2018} in the \emph{decoupling limit} (i.e., simply integrating out the scalar degrees of freedom).
The two above-reported results are perfectly consistent with each other and in agreement with the available most accurate lattice determination, that is, $\chi^{(lattice)}_{QCD} = (75.6 \pm 2.0~ \text{MeV})^4$ \cite{lattice_4},
and also with the results found using the $SU(2)$ chiral perturbation theory (i.e., with $n_l=2$ light flavors) up to the next-to-leading order, $\chi^{(NLO \chi PT_{(2)})}_{QCD} = (75.5 \pm 0.5~ \text{MeV})^4$ \cite{GHVV2016}, and up to the next-to-next-to-leading order, $\chi^{(NNLO \chi PT_{(2)})}_{QCD} = (75.44 \pm 0.34~ \text{MeV})^4$ \cite{GV2019}.

\newpage

\renewcommand{\Large}{\large}

\end{document}